\documentstyle[12pt]{article}

\font\twlgot =eufm10 scaled \magstep1 \font\egtgot =eufm8
\font\sevgot =eufm7

\font\twlmsb =msbm10 scaled \magstep1 \font\egtmsb =msbm8
\font\sevmsb =msbm7

\newfam\gotfam
\def\pgot{\fam\gotfam\twlgot}
\textfont\gotfam\twlgot \scriptfont\gotfam\egtgot
\scriptscriptfont\gotfam\sevgot
\def\got{\protect\pgot}

\newfam\msbfam
\textfont\msbfam\twlmsb \scriptfont\msbfam\egtmsb
\scriptscriptfont\msbfam\sevmsb

\def\pBbb{\relax\ifmmode\expandafter\Bb\else\typeout{You cann't use
Bbb in text mode}\fi}
\def\Bb #1{{\fam\msbfam\relax#1}}

\def\thebibliography#1{\section*{References}\list
  {[\arabic{enumi}]}{\settowidth\labelwidth{#1}\leftmargin\labelwidth
    \advance\leftmargin\labelsep
    \usecounter{enumi}}
    \def\newblock{\hskip .11em plus .33em minus .07em}
    \sloppy\clubpenalty4000\widowpenalty4000
    \sfcode`\.=1000\relax}

\def\op#1{\mathop{\fam0 #1}\limits}

\newcommand{\id}{{\rm Id\,}}

\newcommand{\Ker}{{\rm Ker\,}}
\newcommand{\nm}[1]{\mid {#1}\mid}

\newcommand{\beq}{\begin{equation}}
\newcommand{\eeq}{\end{equation}}
\newcommand{\ben}{\begin{eqnarray}}
\newcommand{\een}{\end{eqnarray}}
\newcommand{\be}{\begin{eqnarray*}}
\newcommand{\ee}{\end{eqnarray*}}
\newcommand{\bea}{\begin{eqalph}}
\newcommand{\eea}{\end{eqalph}}

\newcommand{\cG}{{\got g}}

\newcommand{\cP}{{\cal P}}
\newcommand{\cR}{{\cal R}}
\newcommand{\cL}{{\cal L}}
\newcommand{\cV}{{\cal V}}

\newcommand{\cH}{{\cal H}}
\newcommand{\cF}{{\cal F}}
\newcommand{\cM}{{\cal M}}
\newcommand{\cS}{{\cal S}}

\newcommand{\cN}{{\cal N}}
\newcommand{\ccG}{{\cal G}}
\newcommand{\bL}{{\bf L}}

\newcommand{\al}{\alpha}
\newcommand{\bt}{\beta}
\newcommand{\dl}{\delta}
\newcommand{\la}{\lambda}
\newcommand{\La}{\Lambda}
\newcommand{\f}{\phi}
\newcommand{\om}{\omega}

\newcommand{\m}{\mu}
\newcommand{\n}{\nu}

\newcommand{\G}{\Gamma}

\newcommand{\ve}{\varepsilon}

\newcommand{\vt}{\vartheta}
\newcommand{\si}{\sigma}
\newcommand{\w}{\wedge}

\newcommand{\wh}{\widehat}
\newcommand{\ol}{\overline}
\newcommand{\dr}{\partial}
\newcommand{\ar}{\op\longrightarrow}

\newcommand{\ot}{\otimes}

\newcounter{theorem}
\newcounter{remark}
\newcounter{proposition}
\newcounter{lemma}
\newcounter{corollary}
\newcounter{definition}

\def\theremark{\arabic{remark}}

\def\thedefinition{\arabic{definition}}

\newenvironment{proof}{\noindent {Proof.}}{\medskip}

\newenvironment{rem}{\refstepcounter{remark}\medskip\noindent{Remark
\theremark.}}{\medskip}

\newenvironment{theo}{\refstepcounter{definition} \medskip
\noindent{Theorem \thedefinition.}\it }{\medskip}
\newenvironment{prop}{\refstepcounter{definition} \medskip
\noindent{Proposition \thedefinition.}\it }{\medskip}

\newcommand{\mar}[1]{}

\begin{document}
\hbox{}

\begin{center}

{\large\bf BV QUANTIZATION OF COVARIANT (POLYSYMPLECTIC)
HAMILTONIAN FIELD THEORY}
\bigskip

{\sc D. BASHKIROV}
\medskip

\begin{small}

{\it Department of Theoretical Physics, Physics Faculty, Moscow
State University, \\ 117234 Moscow, Russia}

{\it E-mail: bashkir@phys.msu.ru}

\end{small}
\end{center}

{\parindent=0pt

{\small
\bigskip

Covariant (polysymplectic) Hamiltonian field theory is the
Hamiltonian counterpart of classical Lagrangian field theory. They
are quasi-equivalent in the case of almost-regular Lagrangians.
This work addresses BV quantization of polysymplectic Hamiltonian
field theory. We compare BV quantizations of associated Lagrangian
and polysymplectic Hamiltonian field systems in the case of
almost-regular quadratic Lagrangians.
\bigskip


}

}

\section{Introduction}

The Hamiltonian counterpart of classical first-order Lagrangian
field theory is covariant Hamiltonian formalism which is developed
in the polysymplectic, multisymplectic and Hamilton -- De Donder
variants (see \cite{book,jpa99,ech00,hel,ech04} and references
therein). In order to quantize covariant Hamiltonian field theory,
one usually attempts to construct the multisymplectic
generalization of a Poisson bracket \cite{kanat,for,lop}. We
provide its BV quantization.

Let us consider a field system represented by sections of a fiber
bundle $\pi:Y\to X$ coordinated by $(x^\la,y^i)$. Its
configuration space is the first-order jet manifold $J^1Y$ of $Y$
equipped with the adapted coordinates $(x^\m,y^i,y^i_\m)$,
compatible with the composite fibration \be J^1Y\ar^{\pi^1_0} Y\ar
X. \ee A first-order Lagrangian of fields is defined as a
horizontal density \mar{cmp1}\beq L=\cL\om: J^1Y\to\op\w^nT^*X,
\qquad \om=dx^1\w\cdots dx^n, \qquad n=\dim X, \label{cmp1} \eeq
on the jet manifold $J^1Y$. The corresponding Euler--Lagrange
equations are given by the subset \mar{b327}\beq \dl\cL_i=(\dr_i-
d_\la\dr^\la_i)\cL=0,   \qquad
 d_\la=\dr_\la + y^i_\la\dr_i + y^i_{\la\m}\dr^\m_i, \label{b327}
\eeq of the second-order jet manifold $J^2Y$ of $Y$ coordinated by
$(x^\m,y^i,y^i_\la, y^i_{\la\m})$.

The polysymplectic phase space of a field system on a fiber bundle
$Y\to X$ is the Legendre bundle \mar{00}\beq
\Pi=\op\w^nT^*X\op\ot_YV^*Y\op\ot_YTX=V^*Y\w(\op\w^{n-1}T^*X)
\label{00} \eeq equipped with the holonomic bundle coordinates
$(x^\la,y^i,p^\m_i)$ \cite{book,jpa99}. They are compatible with
the composite fibration \be \Pi\ar^{\pi_Y} Y\ar X. \ee A covariant
Hamiltonian $\cH$ on $\Pi$ (\ref{00}) is defined as a section
$p=-\cH$ of the trivial one-dimensional fiber bundle \mar{N41}\beq
Z_Y=T^*Y\w(\op\w^{n-1}T^*X)\to \Pi, \label{N41} \eeq coordinated
by $(x^\la,y^i,p^\m_i,p)$. This fiber bundle is provided with the
canonical multisymplectic Liouville form \be \Xi= p\om + p^\la_i
dy^i\w\om_\la, \qquad \om_\la=\dr_\la\rfloor\om. \ee The pull-back
of $\Xi$ onto $\Pi$ by a Hamiltonian $\cH$ is a Hamiltonian form
\mar{b418}\beq
 H=\cH^*\Xi_Y= p^\la_i dy^i\w \om_\la -\cH\om  \label{b418}
\eeq
 on $\Pi$. The corresponding covariant Hamilton equations on $\Pi$
are given by the closed submanifold \mar{b4100}\beq
y^i_\la=\dr^i_\la\cH, \qquad  p^\la_{\la i}=-\dr_i\cH
\label{b4100} \eeq of the jet manifold $J^1\Pi$ of $\Pi\to X$. A
covariant Hamiltonian system on $\Pi$ is equivalent to a
particular Lagrangian system on $\Pi$ as follows \cite{jpa99}.

\begin{prop} \label{m11} \mar{m11}
The covariant Hamilton equations (\ref{b4100}) are equivalent to
the Euler--Lagrange equations for the first-order Lagrangian
\mar{m5}\beq L_\cH=h_0(H)=\cL_\cH\om= (p^\la_i y^i_\la-\cH)\om
\label{m5} \eeq on $J^1\Pi$, where $h_0$ is the so called
horizontal projector sending exterior forms on $\Pi$ onto
horizontal exterior forms on $J^1\Pi\to X$ by the rule \be
h_0(dy^i)=y^i_\la dx^\la, \qquad h_0(dp^\m_i)=p^\m_{\la i}dx^\la.
\ee
\end{prop}

This fact motivates us to quantize covariant Hamiltonian field
theory with a Hamiltonian $\cH$ on $\Pi$ as a Lagrangian system
with the Lagrangian $L_\cH$ (\ref{m5}). This Lagrangian system can
be quantized in the framework of familiar perturbation quantum
field theory. If there is no constraint and the matrix \be
\dr^2\cH/\dr p^\m_i\dr p^\nu_j=-\dr^2\cL/\dr p^\m_i\dr p^\nu_j \ee
is positive-definite and nondegenerate on an Euclidean space-time,
this quantization is given by the generating functional
\mar{m2}\beq Z=\cN^{-1}\int\exp\{\int(\cL_\cH +\Lambda +
iJ_iy^i+iJ^i_\m p^\m_i) \om \}\op\prod_x [dp(x)][dy(x)] \label{m2}
\eeq of Euclidean Green functions \cite{sard94}, where $\Lambda$
comes from the normalization condition \be \int
\exp\{\int(\frac12\dr_\m^i\dr_\nu^j\cL_\cH p^\m_i
p^\nu_j+\La)dx\}\op\prod_x[dp(x)]=1. \ee A constrained covariant
Hamiltonian system can be quantized as follows.

Let $i_N:N\to \Pi$ be a closed imbedded subbundle of the Legendre
bundle $\Pi\to Y$ which is regarded as a constraint space of a
covariant Hamiltonian field system with a Hamiltonian $\cH$. Let
$H_N=i^*_NH$ be the pull-back of the Hamiltonian form $H$
(\ref{b418}) onto $N$. This form defines the constrained
Lagrangian \mar{cmp81}\beq L_N=h_0(H_N)=(J^1i_N)^*L_\cH
\label{cmp81} \eeq on the jet manifold $J^1N_L$ of the fiber
bundle $N_L\to X$. The Euler--Lagrange equations for this
Lagrangian are called the constrained Hamilton equations.

\begin{rem} \label{y24} \mar{y24}
In fact, the Lagrangian $L_\cH$ (\ref{m5}) is the pull-back onto
$J^1\Pi$ of the horizontal form $L_\cH$ on the bundle product
$\Pi\op\times_Y J^1Y$ over $Y$ by the canonical map $J^1\Pi\to
\Pi\op\times_Y J^1Y$. Therefore, the constrained Lagrangian $L_N$
(\ref{cmp81}) is simply the restriction of $L_\cH$ to
$N\op\times_Y J^1Y$.
\end{rem}

Let us refer to the following results \cite{jpa99}.

\begin{prop} \label{m10} \mar{m10}
A section $r$ of $\Pi\to X$ is a solution of the covariant
Hamilton equations (\ref{b4100}) iff it satisfies the condition
$r^*(u_\Pi\rfloor dH)= 0$ for any vertical vector field $u_\Pi$ on
$\Pi\to X$.
\end{prop}

\begin{prop} \label{m12} \mar{m12}
A section $r$ of the fiber bundle $N\to X$ is a solution of
constrained Hamilton equations iff it satisfies the condition
$r^*(u_N\rfloor dH)= 0$ for any vertical vector field $u_N$ on
$N\to X$.
\end{prop}

Propositions \ref{m10} and \ref{m12} result in the following.

\begin{prop} \label{y23} \mar{y23}
Any solution of the covariant Hamilton equations (\ref{b4100})
which lives in the constraint manifold $N$ is also a solution of
the constrained Hamilton equations on $N$.
\end{prop}

This fact motivates us to quantize covariant Hamiltonian field
theory on a constraint manifold $N$ as a Lagrangian system with
the pull-back Lagrangian $L_N$ (\ref{cmp81}). Furthermore, a
closed imbedded constraint submanifold $N$ of $\Pi$ admits an open
neighbourhood $U$ which is a fibered manifold $U\to N$. If $\Pi$
is a fibered manifold $\pi_N:\Pi\to N$ over $N$, it is often
convenient to quantize a Lagrangian system on $\Pi$ with the
pull-back Lagrangian $L_\Pi=\pi_N^*L_N$. Since this Lagrangian
possesses gauge symmetries, Batalin--Vilkoviski (henceforth BV)
quantization should be called into play.

Recall that BV quantization \cite{bat,gom} provides the universal
scheme of quantization of gauge-invariant Lagrangian field
systems. Given a classical Lagrangian, it enables one to obtain a
gauged-fixed BRST invariant Lagrangian in the generating
functional of perturbation quantum field theory. However, the BV
quantization scheme does not automatically provide the
path-integral measure.

Here, we apply BV quantization to covariant Hamiltonian systems
associated to Lagrangian field systems with quadratic Lagrangians
\mar{N12}\beq \cL=\frac12 a^{\la\m}_{ij} y^i_\la y^j_\m + b^\la_i
y^i_\la + c, \label{N12} \eeq where $a$, $b$ and $c$ are functions
on $Y$. If the Lagrangian (\ref{N12}) is hyperregular (i.e., the
matrix function $a$ is nondegenerate), there exists a unique
associated Hamiltonian system whose Hamiltonian $\cH$ is quadratic
in momenta $p^\m_i$, and so is the Lagrangian $\cL_\cH$
(\ref{m5}). If the matrix function $a$ is positive-definite on an
Euclidean space-time, the generating functional (\ref{m2}) is a
Gaussian integral of momenta $p^\m_i(x)$. Integrating $Z$ with
respect to $p^\m_i(x)$, one restarts the generating functional of
quantum field theory with the original Lagrangian $\cL$
(\ref{N12}). Using the BV quantization procedure, we generalize
this result to field theories  with almost-regular Lagrangians
$\cL$ (\ref{N12}), e.g., Yang--Mills gauge theory. The key point
is that such a Lagrangian $\cL$ yields constraints and admits
different associated Hamiltonians $\cH$, but all the Lagrangians
$\cL_\cH$ coincide on the constraint manifold and, therefore, we
have a unique constrained Hamiltonian system which is
quasi-equivalent to the original Lagrangian one.

\section{Associated Lagrangian and Hamiltonian systems}

In order to relate classical Lagrangian and covariant Hamiltonian
field theories, let us recall that, besides the Euler--Lagrange
equations, a Lagrangian $L$ (\ref{cmp1}) also yields the Cartan
equations which are given by the subset \mar{b336}\ben && (\ol
y^j_\m- y^j_\m)\dr^\la_i\dr^\m_j\cL=0, \qquad \dr_i\cL-\ol
d_\la\dr_i^\la\cL +(\ol y^j_\m- y^j_\m)\dr_i\dr^\m_j\cL=0,
\label{b336}\\
&& \ol d_\la=\dr_\la +\ol y^i_\la\dr_i +\ol y^i_{\la\m}\dr^\m_i,
\nonumber \een of the repeated jet manifold $J^1J^1Y$ coordinated
by $(x^\m,y^i,y^i_\la,\ol y^i_\la,\ol y^i_{\la\m})$. The jet
prolongation $J^1s$ of any solution $s$ of the Euler--Lagrange
equations (\ref{b327}) is a solution of the Cartan equations
(\ref{b336}). If $\ol s$ is a solution of the Cartan equations and
$\ol s=J^1s$, then $s$ is a solution of the Euler--Lagrange
equations. If a Lagrangian $L$ is regular, the equations
(\ref{b327}) and (\ref{b336}) are equivalent.

Any Lagrangian $L$ (\ref{cmp1}) yields the Legendre map
\mar{m3}\beq \wh L: J^1Y\ar_Y \Pi, \qquad p^\la_i\circ\wh
L=\dr^\la_i\cL, \label{m3} \eeq over $\id Y$ whose image $N_L=\wh
L(J^1Y)$ is called the Lagrangian constraint space. A Lagrangian
$L$ is said to be hyperregular if the Legendre map (\ref{m3}) is a
diffeomorphism. A Lagrangian $L$ is called almost-regular if the
Lagrangian constraint space  is a closed imbedded subbundle
$i_N:N_L\to \Pi$ of the Legendre bundle $\Pi\to Y$ and the
surjection $\wh L:J^1Y\to N_L$ is a fibered manifold possessing
connected fibers. Conversely, any Hamiltonian $\cH$ yields the
Hamiltonian map \mar{415}\beq \wh H: \Pi\ar_Y J^1Y, \qquad
y_\la^i\circ\wh H=\dr^i_\la\cH. \label{415} \eeq

A Hamiltonian $\cH$ on $\Pi$ is said to be associated to a
Lagrangian $L$ on $J^1Y$ if $\cH$ satisfies the relations
\mar{2.30a,b}\ben &&\wh L\circ\wh H\circ \wh L=\wh L, \qquad
p^\m_i=\dr^\m_i\cL (x^\m,y^i,\dr^j_\la\cH), \qquad
(x^\m,y^i,p^\m_i)\in N_L,
\label{2.30a} \\
&&\wh H^*L_\cH=\wh H^*L, \qquad
p^\m_i\dr^i_\m\cH-\cH=\cL(x^\m,y^j,\dr^j_\la\cH), \label{2.30b}
\een where $L_\cH$ is the Lagrangian (\ref{m5}) on $\Pi\op\times_Y
J^1Y$ (see Remark \ref{y24}). If an associated Hamiltonian $\cH$
exists, the Lagrangian constraint space $N_L$ is given by the
coordinate relations (\ref{2.30a}) and $\wh L\circ \wh H$ is a
projector of $\Pi$ onto $N_L$.

For instance, any hyperregular Lagrangian $L$ admits a unique
associated Hamiltonian $\cH$ such that \be \wh H=\wh L^{-1},
\qquad \cH=p^\m_i \wh L^{-1}{}_\m^i -\cL(x^\la, y^i, \wh
L^{-1}{}_\la^i). \ee In this case, any solution $s$ of the
Euler--Lagrange equations (\ref{b327}) defines the solution $r=\wh
L\circ J^1s$ of the covariant Hamilton equations (\ref{b4100}).
Conversely, any solution $r$ of these Hamilton equations yields
the solution $s=\pi_Y\circ r$ of the Euler--Lagrange equations
(\ref{b327}).

A degenerate Lagrangian need not admit an associated Hamiltonian.
If such a Hamiltonian exists, it is not necessarily unique. Let us
restrict our consideration to almost-regular Lagrangians. From the
physical viewpoint, the most of Lagrangian field theories is of
this type. From the mathematical one, this notion of degeneracy is
particularly appropriate for the study of relations between
Lagrangian and covariant Hamiltonian formalisms as follows.

\begin{theo}\label{3.23} \mar{3.23}
Let $L$ be an almost-regular Lagrangian and $\cH$ an associated
Hamiltonian. Let a section $r$ of $\Pi\to X$ be a  solution of the
covariant Hamilton equations (\ref{b4100}) for $\cH$. If $r$ lives
in the Lagrangian constraint manifold $N_L$, then $s=\pi_Y\circ r$
satisfies the Euler--Lagrange equations (\ref{b327}) for $L$,
while $\ol s=\wh H\circ r$ obeys the Cartan equations
(\ref{b336}). Conversely, let $\ol s$ be a solution of the Cartan
equations (\ref{b336}) for $L$. If $\cH$ satisfies the relation
\be \wh H\circ \wh L\circ \ol s=J^1(\pi^1_0\circ\ol s), \ee the
section $r=\wh L\circ \ol s$ of the Legendre bundle $\Pi\to X$ is
a solution of the covariant Hamilton equations (\ref{b4100}) for
$\cH$. If $\ol s=J^1s$, we obtain the relation between solutions
the Euler--Lagrange equations and the covariant Hamilton ones.
\end{theo}

By virtue of Theorem \ref{3.23}, one need a set of different
associated Hamiltonians in order to recover all solutions of the
Euler--Lagrange and Cartan equations for an almost-regular
Lagrangian $L$. This ambiguity can be overcome as follows.

\begin{prop} \label{m0} \mar{m0}
Let $\cH$, $\cH'$ be two different Hamiltonians associated to an
almost-regular Lagrangian $L$. Let $H$, $H'$ be the corresponding
Hamiltonian forms (\ref{b418}). Their pull-backs $i^*_NH$ and
$i^*_NH'$ onto the Lagrangian constraint manifold $N_L$ coincide.
\end{prop}

It follows that, if an almost-regular Lagrangian admits associated
Hamiltonians $\cH$, it defines a unique constrained Hamiltonian
form $H_N=i^*_NH$ on the Lagrangian constraint manifold $N_L$ and
a unique constrained Lagrangian $L_N=h_0(H_N)$ (\ref{cmp81}) on
the jet manifold $J^1N_L$ of the fiber bundle $N_L\to X$. Basing
on Proposition \ref{y23} and Theorem \ref{3.23}, one can prove the
following.

\begin{theo}\label{3.01} \mar{3.01} Let an almost-regular Lagrangian
$L$ admit associated Hamiltonians. A section $\ol s$ of the jet
bundle $J^1Y\to X$ is a solution of the Cartan equations for $L$
iff $\wh L\circ \ol s$ is a solution of  the constrained Hamilton
equations. In particular, any solution $r$ of the constrained
Hamilton equations provides the solution $\ol s=\wh H\circ r$ of
the Cartan equations.
\end{theo}

Theorem \ref{3.01}  shows that the constrained Hamilton equations
and the Cartan equations are quasi-equivalent. Thus, one can
associate to an almost-regular Lagrangian $L$ (\ref{cmp1}) a
unique constrained Lagrangian system on the constraint Lagrangian
manifold $N_L$ (\ref{2.30a}). Let us compare quantizations of
these Lagrangian systems on $Y$ and $N_L\subset \Pi$ in the case
of an almost-regular quadratic Lagrangian $L$.

\section{Quadratic degenerate systems}

Given a fiber bundle $Y\to X$, let us consider a  quadratic
Lagrangian $L$ (\ref{N12}), where $a$, $b$ and $c$ are local
functions on $Y$. This property is coordinate-independent since
$J^1Y\to Y$ is an affine bundle modeled over the vector bundle
$T^*X\op\ot_Y VY$. The associated Legendre map (\ref{m3}) reads
\mar{N13}\beq p^\la_i\circ\wh L= a^{\la\m}_{ij} y^j_\m +b^\la_i.
\label{N13} \eeq

Let a Lagrangian $L$ (\ref{N12}) be almost-regular, i.e., the
matrix function $a$ is a linear bundle morphism \mar{m38}\beq a:
T^*X\op\ot_Y VY\to \Pi, \qquad p^\la_i=a^{\la\m}_{ij} \ol y^j_\m,
\label{m38} \eeq of constant rank, where $(x^\la,y^i,\ol y^i_\la)$
are bundle coordinates on $T^*X\op\ot_Y VY$. Then the Lagrangian
constraint space $N_L$ (\ref{N13}) is an affine subbundle of the
Legendre bundle $\Pi\to Y$. Hence, $N_L\to Y$ has a global
section. For the sake of simplicity, let us assume that it is the
canonical zero section $\wh 0(Y)$ of $\Pi\to Y$. The kernel of the
Legendre map (\ref{N13})  is also an affine subbundle of the
affine jet bundle $J^1Y\to Y$. Therefore, it admits a global
section \mar{N16}\beq \G: Y\to \Ker\wh L\subset J^1Y, \qquad
a^{\la\m}_{ij}\G^j_\m + b^\la_i =0,  \label{N16} \eeq which is a
connection on $Y\to X$. With such a connection, the Lagrangian
(\ref{N12}) is brought into the form \mar{y47}\beq \cL=\frac12
a^{\la\m}_{ij} (y^i_\la -\G^i_\la)(y^j_\m-\G^j_\m) +c'.
\label{y47} \eeq

Let us refer to the following theorems \cite{book,jpa99}.

\begin{theo}\label{04.2}  There exists a linear bundle
morphism \mar{N17}\beq \si: \Pi\op\to_Y T^*X\op\otimes_YVY, \qquad
\ol y^i_\la\circ\si =\si^{ij}_{\la\m}p^\m_j, \label{N17} \eeq such
that \mar{N45}\beq a\circ\si\circ a=a, \qquad
a^{\la\mu}_{ij}\si^{jk}_{\mu\al}a^{\al\nu}_{kb}=a^{\la\nu}_{ib}.
\label{N45} \eeq
\end{theo}

The equalities (\ref{N16}) and (\ref{N45}) give the relation
$(a\circ\si_0)^{\la j}_{i\m} b^\m_j=b^\la_i$. Note that the
morphism $\si$ (\ref{N17}) is not unique, but it falls into the
sum $\si=\si_0+\si_1$ such that \mar{N21}\beq \si_0\circ a\circ
\si_0=\si_0, \qquad a\circ\si_1=\si_1\circ a=0, \label{N21} \eeq
where $\si_0$ is uniquely defined.

\begin{theo} \label{m15} \mar{m15}
There are the splittings \mar{N18,20}\ben && J^1Y=\Ker\wh
L\op\oplus_Y{\rm Im}(\si_0\circ
\wh L), \label{N18} \\
&& y^i_\la=\cS^i_\la+\cF^i_\la= [y^i_\la -\si_0{}^{ik}_{\la\al}
(a^{\al\m}_{kj}y^j_\m + b^\al_k)]+
[\si_0{}^{ik}_{\la\al} (a^{\al\m}_{kj}y^j_\m + b^\al_k)], \label{N18'}\\
&& \Pi=\Ker\si_0 \op\oplus_Y N_L, \label{N20} \\
&& p^\la_i = \cR^\la_i+\cP^\la_i= [p^\la_i -
a^{\la\m}_{ij}\si_0{}^{jk}_{\m\al}p^\al_k] +
[a^{\la\m}_{ij}\si_0{}^{jk}_{\m\al}p^\al_k]. \label{N20'} \een
\end{theo}

The relations (\ref{N21}) lead to the equalities \mar{m25}\beq
\si_0{}^{jk}_{\m\al}\cR^\al_k=0, \qquad
\si_1{}^{jk}_{\m\al}\cP^\al_k=0, \qquad \cR^\la_i\cF^i_\la=0.
\label{m25} \eeq

By virtue of the equalities (\ref{N21}) and the relation
\mar{y75}\beq \cF^i_\m=(\si_0\circ a)^{i\la}_{\m
j}(y^j_\la-\G^j_\la), \label{y75} \eeq the Lagrangian (\ref{N12})
takes the form \mar{cmp31}\beq L=\cL\om, \qquad \cL=\frac12
a^{\la\m}_{ij}\cF^i_\la\cF^j_\m +c'. \label{cmp31} \eeq One can
show that this Lagrangian admits a set of associated Hamiltonians
\mar{N22}\beq \cH_\G=(\cR^\la_i+\cP^\la_i)\G^i_\la +\frac12
\si_0{}^{ij}_{\la\m}\cP^\la_i\cP^\m_j
+\frac12\si_1{}^{ij}_{\la\m}\cR^\la_i\cR^\m_j -c'\label{N22} \eeq
indexed by connections $\G$ (\ref{N16}). Accordingly, the
Lagrangian constraint manifold (\ref{N13}) is characterized by the
equalities \mar{bv1}\beq \cR^\la_i=p^\la_i -
a^{\la\m}_{ij}\si_0{}^{jk}_{\m\al}p^\al_k=0. \label{bv1} \eeq

Given a Hamiltonian $\cH_\G$, the corresponding Lagrangian
(\ref{m5}) on $\Pi\op\times_Y J^1Y$ (see Remark \ref{y24}) reads
\mar{m16}\beq \cL_{\cH_\G}=\cR^\la_i(\cS^i_\la-\G^i_\la)
+\cP^\la_i\cF_\la^i -\frac12\si_0{}^{ij}_{\la\m}\cP^\la_i\cP^\m_j
- \frac12\si_1{}^{ij}_{\la\m} \cR^\la_i \cR^\m_j+ c'. \label{m16}
\eeq Its restriction (\ref{cmp81}) to the constraint manifold
$N_L\op\times_YJ^1Y$ is \mar{bv2}\beq L_N=\cL_N\om, \qquad
\cL_N=\cP^\la_i\cF_\la^i
-\frac12\si_0{}^{ij}_{\la\m}\cP^\la_i\cP^\m_j + c'. \label{bv2}
\eeq It is independent of the choice of a Hamiltonian (\ref{N22}).

The Hamiltonian $\cH_\G$ yields the Hamiltonian map $\wh H_\G$
(\ref{415}) and the projector \be T=\wh L\circ \wh H_\G,\qquad
p^\la_i\circ T=T^{\la j}_{i
\m}p^\m_j=a^{\la\nu}_{ik}\si_0{}^{kj}_{\nu\m}
p^\m_j=\cP^\la_i,\label{m30} \ee of $\Pi$ onto its summand $N_L$
in the decomposition (\ref{N20}). It is a linear morphism over
$\id Y$. Therefore, $T:\Pi\to N_L$ is a vector bundle. We aim to
quantize the pull-back \mar{m32}\beq L_\Pi=T^*L_N=\cL_\Pi\om,
\qquad \cL_\Pi=\cP^\la_i\cF_\la^i
-\frac12\si_0{}^{ij}_{\la\m}\cP^\la_i \cP^\m_j + c', \label{m32}
\eeq of the constrained Lagrangian $L_N$ (\ref{bv2}) onto
$\Pi\op\times_Y J^1Y$.

Note that the splittings (\ref{N18}) and (\ref{N20}) result from
the splitting of the vector bundle \be T^*X\op\ot_Y VY=\Ker
a\op\oplus_Y E, \ee which can be provided with the adapted
coordinates $(\ol y^a,\ol y^A)$ such that $a$ (\ref{m38}) is
brought into a diagonal matrix with nonvanishing components
$a_{AA}$. Then the Legendre bundle $\Pi\to Y$ (\ref{00}) is
endowed with the dual (nonholonomic) coordinates $(p_a,p_A)$ where
$p_A$ are coordinates on the Lagrangian constraint manifold $N_L$,
given by the equalities $p_a=0$. Written relative to these
coordinates, $\si_0$ becomes the diagonal matrix \mar{m39}\beq
\si_0^{AA}=(a_{AA})^{-1}, \qquad \si_0^{aa}=0, \label{m39} \eeq
while $\si_1^{Aa}=\si_1^{AB}=0$.  Let us write \mar{m41}\beq
p_a=M_a{}^i_\la p^\la_i, \qquad p_A=M_A{}^i_\la p^\la_i,
\label{m41} \eeq where $M$ are the matrix functions on $Y$ obeying
the relations \mar{y30}\beq M_a{}^i_\la a^{\la\m}_{ij}=0, \quad
(M^{-1})^a{}_i^\la\si_0{}^{ij}_{\la\m}=0, \quad
M_A{}^i_\la(a\circ\si_0)^{\la j}_{i\m}= M_A{}^j_\m, \quad
(M^{-1})^A{}_j^\m M_A{}^i_\la=
a_{jk}^{\m\nu}\si_0{}^{ki}_{\nu\la}. \label{y30} \eeq

\section{Gauge symmetries}

The Lagrangian $L_\Pi$ (\ref{m32}) possesses gauge symmetries. By
gauge transformations are meant automorphisms $\Phi$ of the
composite fiber bundle $\Pi\to Y\to X$ over bundle automorphisms
$\f$ of $Y\to X$ over $\id X$. Such an automorphism $\Phi$ gives
rise to the automorphism $(\Phi, J^1\f)$ of the composite fiber
bundle \be \Pi\op\times_Y J^1Y\to Y\to X. \ee An automorphism
$\Phi$ is said to be a gauge symmetry of the Lagrangian $L_\Pi$ if
$(\Phi,J^1\f)^*L_\Pi=L_\Pi$. If the Lagrangian (\ref{N12}) is
degenerate, the group $G$ of gauge symmetries of the Lagrangian
$L_\Pi$ (\ref{m32}) is never trivial. Indeed, any vertical
automorphism of the vector bundle $\Ker \si_0\to Y$ in the
decomposition (\ref{N20}) is obviously a gauge symmetry of the
Lagrangian $L_\Pi$ (\ref{m32}). The gauge group $G$ acts on the
space $\Pi(X)$ of sections of the Legendre bundle $\Pi\to X$. For
the purpose of quantization, it suffices to consider a subgroup
$\ccG$ of $G$  which acts freely on $\Pi(X)$ and satisfies the
relation $\Pi(X)/\ccG=\Pi(X)/G$. Moreover, we need one-parameter
subgroups of $\ccG$. Their infinitesimal generators are
represented by projectable vector fields \mar{y20}\beq
u_\Pi=u^i(x^\m,y^j)\dr_i + u^\la_i(x^\m,y^j,p^\m_j)\dr^i_\la
\label{y20} \eeq on the Legendre bundle $\Pi\to Y$ which give rise
to the vector fields \mar{m47}\beq \ol u=u^i\dr_i +
u_i^\la\dr^i_\la +d_\la u^i\dr_i^\la, \qquad d_\la=\dr_\la
+y_\la^i\dr_i, \label{m47} \eeq on $\Pi\op\times_Y J^1Y$. A
Lagrangian $L_\Pi$ is invariant under a one-parameter group of
gauge transformations iff its Lie derivative \be \bL_{\ol
u}L_\Pi=\ol u(\cL_\Pi)\om \ee along the infinitesimal generator
$\ol u$ (\ref{m47}) of this group vanishes.

Since linear and affine spaces of fields are only quantized, let
us assume that $Y\to X$ is an affine bundle modeled over a vector
bundle $\ol Y\to X$ (e.g., $\ol Y=Y$ if $Y$ is a vector bundle).
In this case, the vertical tangent and cotangent bundles $VY$ and
$V^*Y$ of $Y\to X$ are canonically isomorphic to the products
\mar{y70}\beq VY=Y\op\times_X \ol Y, \qquad V^*Y=Y\op\times_X \ol
Y^*. \label{y70} \eeq Accordingly, the Legendre bundle $\Pi\to X$
(\ref{00}) is isomorphic to the product \be \Pi=Y\op\times_X (\ol
Y^*\op\ot_X\op\w^n T^*X\op\ot_X TX) \ee such that transition
functions of coordinates $p^\la_i$ are independent of $y$. Then
the splitting (\ref{N20}) takes the form \mar{y26}\beq
\Pi=Y\op\times_X(\ol{\Ker \si_0}\op\oplus_X\ol N_L), \label{y26}
\eeq where $\ol{\Ker \si_0}$ and $\ol N_L$ are fiber bundles over
$X$ such that $\Ker \si_0=\pi^*\ol{\Ker \si_0}$ and $N_L=\pi^*\ol
N_L$ are their pull-backs onto $Y$. The splitting (\ref{y26})
keeps the coordinate form (\ref{N20'}). Due to the splittings
(\ref{y70}) and (\ref{y26}), one can choose the coordinates $p_a$
on $\Ker\si_0\to Y$ such that $\si_1$ becomes a diagonal matrix
with nonzero positive entities $\si_1^{ab}=\dl^{ab}\cV^{-1}$,
where $\cV\om$ is a volume form on $X$. In this case, the matrix
functions $(M^{-1})^a{}_j^\m$ (\ref{m41}) are independent of
$y^j$.

The splittings (\ref{N18}) and (\ref{y26}) lead to the
decomposition \mar{y25}\beq \Pi\op\times_Y J^1Y=(\ol{\Ker
\si_0}\op\oplus_X\ol N_L)\op\times_Y (\Ker\wh L\op\oplus_Y {\rm
Im}(\si_0\circ\wh L)). \label{y25} \eeq In view of this
decomposition, let us define the gauge group $\ccG$ as a direct
product of the additive group $\ccG_\Pi$ of sections of the vector
bundle $\ol{\Ker\si_0}\to X$ and some group $\ccG_Y$ of gauge
symmetries of the Lagrangian $L$ (\ref{N12}). The infinitesimal
generators of the group $\ccG_\Pi$ are vector fields \mar{y27}\beq
u_\Pi=\xi_a (M^{-1})^a{}_i^\la\dr^i_\la, \label{y27} \eeq
parameterized by components $\xi_a$ of sections of $\ol{\Ker
\si_0}\to X$ with respect to the coordinates $p_a$ (\ref{m41}).
The vector fields (\ref{y27}) mutually commute. Their lift
(\ref{m47}) onto $\Pi\op\times_Y J^1Y$ keeps the coordinate form
\mar{y27'}\beq \ol u_\Pi=\xi_a (M^{-1})^a{}_i^\la\dr^i_\la
\label{y27'} \eeq (\ref{y27}). By virtue of the relations
(\ref{y30}), the Lie derivatives of $L_\Pi$ along vector fields
(\ref{y27'}) vanish.

Though it may happen that the Lagrangian $L$ (\ref{N12}) does not
possess gauge symmetries, let us assume that it is invariant under
some group $\ccG_Y$ of vertical automorphisms of the fiber bundle
$Y\to X$. The infinitesimal generators of one-parameter subgroups
of $\ccG_Y$ are represented by vertical vector fields $u=u^i\dr_i$
on $Y\to X$ which give rise to the vector fields \mar{m47'}\beq
J^1u=u^i\dr_i +d_\la u^i\dr_i^\la, \qquad d_\la=\dr_\la
+y_\la^i\dr_i, \label{m47'} \eeq on $J^1Y$. The Lie derivatives
$\bL_{J^1u}L$ of the Lagrangian $L$ (\ref{cmp31}) along the vector
fields (\ref{m47}) vanish, i.e., \mar{m49}\beq (u^i\dr_i +d_\la
u^i\dr_i^\la)\cL=0. \label{m49} \eeq

In order to study the invariance condition (\ref{m49}), let us
consider the Lagrangian (\ref{N12}) written in the form
(\ref{y47}). Since \mar{y50}\beq
J^1u(y^i_\la-\G^i_\la)=\dr_ku^i(y^k_\la-\G^k_\la), \label{y50}
\eeq one easily obtains from the equality (\ref{m49}) that
\mar{y45}\beq u^k\dr_k a^{\la\m}_{ij}+ \dr_iu^k a^{\la\m}_{kj} +
a^{\la\m}_{ik}\dr_ju^k =0. \label{y45} \eeq It follows that the
summands of the Lagrangian $L$ (\ref{y47}) and, consequently, the
summands of the Lagrangian (\ref{cmp31}) are separately
gauge-invariant, i.e., \mar{y31}\beq
J^1u(a^{\la\m}_{ij}\cF^i_\la\cF^j_\m)=0, \qquad
J^1u(c')=u^k\dr_kc'=0. \label{y31} \eeq The equalities
(\ref{y75}), (\ref{y50}) and (\ref{y45}) give the transformation
law \mar{b1}\beq J^1u(a^{\la\m}_{ij}\cF^j_\m)=-\dr_i u^k
a^{\la\m}_{kj}\cF^j_\m. \label{b1} \eeq The  relations (\ref{N21})
and (\ref{y45}) lead to the equality \mar{y53}\beq
a^{\la\mu}_{ij}[u^k\dr_k\si_0{}^{jn}_{\mu\al} -\dr_ku^j
\si_0{}^{kn}_{\mu\al} - \si_0{}^{jk}_{\mu\al}\dr_ku^n
]a^{\al\nu}_{nb}=0. \label{y53} \eeq

Now, let us compare gauge symmetries of the Lagrangian $L$
(\ref{cmp31}) and the Lagrangian $L_N$ (\ref{bv2}). Given the
Legendre map $\wh L$ (\ref{N13}) and the tangent morphism \be T\wh
L: TJ^1Y\to TN_L, \qquad \dot p_A=(\dot y^i\dr_i +\dot
y^k_\nu\dr^\nu_k) (M_A{}^i_\la a^{\la\m}_{ij}\cF^j_\m), \ee let us
consider the map \mar{y79}\ben && T\wh L\circ J^1u: J^1Y\ni
(x^\la,y^i,y^i_\la) \mapsto u^i\dr_i + (u^k\dr_k +\dr_\nu
u^k\dr^\nu_k) (M_A{}^i_\la
a^{\la\m}_{ij}\cF^j_\m)\dr^A= \label{y79} \\
&& \qquad u^i\dr_i + [u^k\dr_k(M_A{}^i_\la) a^{\la\m}_{ij}\cF^j_\m
+M_A{}^i_\la J^1u
(a^{\la\m}_{ij}\cF^j_\m)]\dr^A= \nonumber\\
&& \qquad  u^i\dr_i + [u^k\dr_k(M_A{}^i_\la)
a^{\la\m}_{ij}\cF^j_\m -M_A{}^i_\la \dr_iu^k
a^{\la\m}_{kj}\cF^j_\m]\dr^A=\nonumber \\
&& \qquad u^i\dr_i + [u^k\dr_k(a\circ\si_0)^{\m
i}_{j\la}\cP^\la_i- (a\circ\si_0)^{\m i}_{j\la}\dr_i
u^k\cP^\la_k]\dr^j_\m\in TN_L, \nonumber \een where the relations
(\ref{y30}) and (\ref{b1}) have been used. Let us assign to a
point $(x^\la,y^i,\cP^\la_i)\in N_L$ some point \mar{y81}\beq
(x^\la,y^i,y^i_\la) \in \wh L^{-1}(x^\la,y^i,\cP^\la_i)
\label{y81} \eeq
 and then the image
of the point (\ref{y81}) under the morphism (\ref{y79}). We obtain
the map \mar{y82}\beq v_N:(x^\la,y^i,\cP^\la_i) \to u^i\dr_i +
[u^k\dr_k(a\circ\si_0)^{\m i}_{j\la}\cP^\la_i- (a\circ\si_0)^{\m
i}_{j\la}\dr_i u^k\cP^\la_k]\dr^j_\m \label{y82} \eeq which is
independent of the choice of a point (\ref{y81}). Therefore, it is
a vector field on the Lagrangian constraint manifold $N_L$. This
vector field gives rise to the vector field \mar{y83}\beq \ol
v_N=u^i\dr_i + [u^k\dr_k(a\circ\si_0)^{\m i}_{j\la}\cP^\la_i-
(a\circ\si_0)^{\m i}_{j\la}\dr_i u^k\cP^\la_k]\dr^j_\m + d_\la
u^i\dr^\la_i \label{y83} \eeq on $N_L\op\times_Y J^1Y$.

\begin{prop} \label{y84} \mar{y84}
The Lie derivative $\bL_{\ol v_N} L_N$ of the Lagrangian $L_N$
(\ref{bv2}) along the vector field $\ol v_N$ (\ref{y83}) vanishes.
\end{prop}

\begin{proof}
One can show that \mar{y87}\beq v_N(\cP^\la_i)=-\dr_iu^k\cP^\la_k
\label{y87} \eeq on the constraint manifold $\cR^\la_i=0$. Then
the invariance condition $\ol v_N(\cL_N)=0$ falls into the three
equalities \mar{y85}\beq \ol
v_N(\si_0{}^{ij}_{\la\m}\cP^\la_i\cP^\m_j)=0, \qquad \ol
v_N(\cP^\la_i\cF^i_\la)=0, \qquad \ol v_N(c')=0. \label{y85} \eeq
The latter is exactly the second equality (\ref{y31}). The first
equality (\ref{y85}) is satisfied due to the relations (\ref{y53})
and (\ref{y87}). The second one takes the form \mar{y90}\beq \ol
v_N(\cP^\la_i(y^i_\la-\G^i_\la))=0. \label{y90} \eeq It holds
owing to the relations (\ref{y50}) and (\ref{y87}).
\end{proof}

Thus, the gauge invariance of the Lagrangian $L$ (\ref{cmp31})
implies that of the Lagrangian $L_N$ (\ref{bv2}). Turn now to
gauge symmetries of the pull-back $L_\Pi$ (\ref{m32}) of the
Lagrangian $L_N$ onto $\Pi\op\times_Y J^1Y$.

Due to the splitting (\ref{y26}), the vector field $v_N$
(\ref{y82}) on $N_L$ can be extended onto $\Pi$ by putting it zero
on $\Ker \si_0$. It keeps the coordinate form \mar{y88}\beq
v_\Pi=u^i\dr_i + [u^k\dr_k(a\circ\si_0)^{\m i}_{j\la}\cP^\la_i-
(a\circ\si_0)^{\m i}_{j\la}\dr_i u^k\cP^\la_k]\dr^j_\m,
\label{y88} \eeq but the transformation law (\ref{y87}) is
modified as \mar{y95}\beq v_\Pi(\cP^\la_i)=
u^k\dr_k(a\circ\si_0)^{\la j}_{i\m}\cP^\m_j -\dr_iu^k\cP^\la_k.
\label{y95} \eeq As a consequence, the invariance condition
(\ref{y90}) does not hold, and the Lagrangian $L_\Pi$ fails to be
gauge-invariant in general.

In order to overcome this difficulty, let us assume that the gauge
group $\ccG_Y$ preserves the splitting (\ref{N18}), i.e., its
infinitesimal generators $u$ obey the condition \mar{y49'}\beq
u^k\dr_k(\si_0{}^{im}_{\la\nu}a^{\nu\m}_{mj})+
\si_0{}^{im}_{\la\nu}a^{\nu\m}_{mk}\dr_ju^k -
\dr_ku^i\si_0{}^{km}_{\la\nu}a^{\nu\m}_{mj} =0. \label{y49'} \eeq
The relations (\ref{y50}) and (\ref{y49'}) lead to the
transformation law \mar{m53}\beq J^1u(\cF^i_\m)=\dr_ju^i\cF^j_\m.
\label{m53} \eeq Since $\cS^i_\la=y^i_\la-\cF^i_\la$, we also
obtain \mar{m54}\beq J^1u(\cS^i_\m)=d_\m u^i-\dr_ju^i\cF^j_\m=
d_\m u^i-\dr_ju^i(y^j_\m-\cS^j_\m) =\dr_\m u^i +\dr_ju^i\cS^j_\m.
\label{m54} \eeq A glance at this relation shows that the gauge
group $\ccG_Y$ acts freely on the space of sections $\cS(x)$ of
the fiber bundle $\Ker\wh L\to Y$ in the splitting (\ref{N18}).
Then some combinations $b^r{}_i^\m(x)\cS^i_\m$ of $\cS^i_\m$ can
be used as the gauge-fixing condition \mar{y1}\beq
b^r{}_i^\m\cS^i_\m(x)=\al^r(x), \label{y1} \eeq similar to the
generalized Lorentz gauge in Yang--Mills gauge theory.

By virtue of the condition (\ref{y49'}), the vector field $v_\Pi$
(\ref{y88}) takes the form \mar{y96}\beq v_\Pi=u^i\dr_i -\dr_i u^k
\cP^\la_k\dr^i_\la. \label{y96} \eeq However, the transformation
law (\ref{y95}) holds and the Lagrangian $L_\Pi$ fails to be
gauge-invariant. Therefore, let us consider a different vector
field on $\Pi$ projected onto the vector field $v_N$ (\ref{y82})
on $N_L$.

Any vertical vector field $u$ on $Y\to X$ gives rise to the vector
field \mar{y41}\beq u_\Pi=u^i\dr_i - \dr_ju^i p^\la_i\dr^j_\la
\label{y41} \eeq on the Legendre bundle $\Pi$ and to the vector
field \mar{y40}\beq \ol u_\Pi=u^i\dr_i - \dr_ju^i p^\la_i\dr^j_\la
+d_\la u^i\dr^\la_i \label{y40} \eeq on $\Pi\op\times_Y J^1Y$. The
vector field (\ref{y41}) as like as the vector field (\ref{y96})
is projected onto the vector field $v_N$ (\ref{y82}). We have \be
u_\Pi-v_\Pi=-\dr_j u^i \cR^\la_i\dr^j_\la. \ee

\begin{prop} \label{y41'} \mar{y41'}
If the condition (\ref{y49'}) holds, the vector field $u_\Pi$
(\ref{y41}) is an infinitesimal gauge symmetry of the Lagrangian
$L_\Pi$ (\ref{m32}) iff $u$ is an infinitesimal gauge symmetry of
the Lagrangian $L$ (\ref{cmp31}).
\end{prop}

\begin{proof}
Due to the condition (\ref{y49'}), the infinitesimal gauge
symmetry $u_\Pi$ (\ref{y41}) preserves the splitting (\ref{N20}),
i.e., \mar{y55}\beq \ol u_\Pi(\cP^\la_i)=-\dr_i u^k\cP^\la_k,
\qquad \ol u(\cR^\la_i)=-\dr_i u^k\cR^\la_k. \eeq We have the
gauge invariance condition \mar{m49'}\beq (u^i\dr_i-\dr_j
u^ip^\la_i\dr^j_\la +d_\la u^i\dr_i^\la)\cL_\Pi=0. \label{m49'}
\eeq It is readily observed that the first and third terms of the
Lagrangian $L_\Pi$ are separately gauge-invariant due to the
relations (\ref{y31}) and (\ref{m53}). Its second term is
gauge-invariant owing to the equality (\ref{y53}). Conversely, let
the invariance condition (\ref{m49'}) holds. It falls into the
independent equalities \mar{m50}\beq \ol
u_\Pi(\si_0{}^{ij}_{\la\m}p^\la_i p^\m_j) =0, \qquad
 \ol u_\Pi(p^\la_i\cF_\la^i)=0, \qquad
u^i\dr_i c'=0, \label{m50} \eeq i.e., the Lagrangian $L_\Pi$ is
gauge-invariant iff its three summands are separately
gauge-invariant. One obtains at once from the second condition
(\ref{m50}) that the quantity $\cF$ is transformed as the dual of
momenta $p$. Then the first condition (\ref{m50}) shows that the
quantity $\si_0p$ is transformed by the same law as $\cF$. It
follows that the term $a\cF\cF$ in the Lagrangian $L$
(\ref{cmp31}) is transformed as $a(\si_0 p)(\si_0 p)=\si_0pp$,
i.e., is gauge-invariant. Then this Lagrangian is gauge-invariant
due to the third equality (\ref{m50}).
\end{proof}

Though vector fields $u_\Pi$ (\ref{y41}) are infinitesimal
generators of gauge symmetries of the Lagrangian $L_\Pi$
(\ref{m32}) in accordance with Proposition \ref{y41'}, they are
not infinitesimal generators of the gauge group $\ccG_Y$ because
they fail to commute with the vector fields (\ref{y27}). Moreover,
the system of vector fields (\ref{y27}) and (\ref{y41}) is not
closed with respect to the Lie bracket, and this fact makes the BV
quantization procedure rather complicated.

Therefore, let us return to the vector fields $v_\Pi$ (\ref{y96}),
but require that \mar{y99}\beq u^k\dr_ka^{\nu\m}_{mj}=0.
\label{y99} \eeq For instance, this condition is always satisfied
if the matrix function $a$ in the Lagrangian (\ref{N12}) is
independent of $y^i$. This is the standard case of perturbation
quantum field theory, e.g., quantum gauge theory.

If $u$ obeys the condition (\ref{y99}), the transformation law
(\ref{y95}) comes to \mar{y100}\beq v_\Pi(\cP^\la_i)=
-\dr_iu^k\cP^\la_k. \label{y100} \eeq Due to this transformation
law, the vector fields (\ref{y96}) become infinitesimal generators
of gauge symmetries of the Lagrangian $L_\Pi$ (\ref{m32}).
Moreover, if the condition (\ref{y99}) holds, vector field
(\ref{y96}) commute with the infinitesimal generators (\ref{y27})
of the gauge group $\ccG_\Pi$. Therefore, they are infinitesimal
generators of the gauge group $\ccG_Y$ acting on $\Pi\op\times_Y
J^1Y$. Thus, the vector fields $u_\Pi$ (\ref{y27}) and $v_\Pi$
(\ref{y96}) are infinitesimal generators of one-parameter groups
of gauge symmetries of the Lagrangian $L_\Pi$ (\ref{m32}) which we
aim to quantize.

Let us further assume that the gauge group $\ccG_Y$ of the
Lagrangian $L$ (\ref{N12}) is indexed by $m$ parameter functions
$\xi^r(x)$ such that the components $u^i(x^\la,y^j,\xi^r)$ of its
infinitesimal generators $u$ are linear first order differential
operators \mar{m48}\beq u^i(x^\la,y^j,\xi^r)=u_r^i(x^\la,y^j)\xi^r
+u_r^{i\m}(x^\la,y^j)\dr_\m\xi^r \label{m48} \eeq on the space of
parameters $\xi^r(x)$ and the vector fields $u(\xi^r)$ (\ref{y20})
satisfy the commutation relations \be
[u(\xi^q),u(\xi'^p)]=u(c^r_{pq}\xi'^p\xi^q), \ee where $c^r_{pq}$
are structure constants.

\section{BV quantization}

Turn now to the BV quantization of the Lagrangian $\cL_\Pi$
(\ref{m32}) on $\Pi\op\times_Y J^1Y$ whose infinitesimal gauge
symmetries are represented by vector field \mar{y101}\beq \ol
u=u^i\dr_i + (\xi_a(M^{-1})^a{}_i^\la  -\dr_i u^k
\cP^\la_k)\dr^i_\la +d_\la u^i\dr^\la_i, \label{y101} \eeq where
$u$ are vector fields (\ref{m48}) depending on parameters $\xi^r$.

We follow the quantization procedure in \cite{bat,gom}
reformulated in the jet terms \cite{barn,bran01}. Note that odd
fields $C^r$ can be introduced as the basis for a graded manifold
determined by the dual $E^*$ of a vector space $E\to X$
coordinated by $(x^\la,e^r)$. Then the $k$-order jets
$C^r_{\la_k\ldots\la_1}$ are defined as the basis for a graded
manifold determined by the dual of the $k$-order jet bundle
$J^kE\to X$, which is a vector bundle \cite{mpl}. The BV
quantization procedure falls into the two steps. At first, one
obtains a proper solution of the classical master equation and,
afterwards, the gauge-fixed BRST invariant Lagrangian is
constructed.

Let the number $m$ of parameters of the gauge group $\ccG$ do not
exceed the fiber dimension of $\Ker\wh L\to Y$. Then we can follow
the standard BV quantization procedure for irreducible gauge
theories in \cite{gom}.

Firstly, one should introduce odd ghosts $C^r$, $C_a$ of ghost
number 1 together with odd antifields $y^*_i$, $p^{*i}_\la$ of
ghost number $-1$ and even antifields $C^*_r$, $C^{*a}$ of ghost
number $-2$. Then a proper solution of the classical master
equation reads \mar{y0}\beq \cL_{\rm PS}=\cL_\Pi + y^*_iu^i_C +
p^{*i}_\la u_{Ci}^\la -\frac12c^r_{pq}C^*_rC^pC^q, \label{y0} \eeq
where $u_C$ is the vector field \mar{y7}\beq u_C^i=u_r^iC^r
+u_r^{i\m}C^r_\m, \qquad u_{Ci}^\la=C_a(M^{-1})^a{}_i^\la -\dr_i
u_C^k\cP^\la_k \label{y7} \eeq obtained from the vector field
(\ref{y101}) by replacement of parameter functions $\xi_a$,
$\xi^r$ and their derivatives $\dr_\m\xi^r$ with the ghosts $C_a$,
$C^r$ and their jets $C^r_\m$.

Secondly, one introduces the gauge-fixing density depending on
fields $y^i$, $p^\la_i$ ghosts $C^r$, $C_a$ and additional
auxiliary fields, which are odd fields $\ol C_r$, $\ol C^a$ of
ghost number $-1$ and even fields $B_r$, $B^a$ of zero ghost
number. Passing to the Euclidean space-time, this gauge-fixing
density reads \mar{y2}\beq \Psi=\ol C_p(\frac{i}{2}  h^{pr}B_r
+b^p{}_i^\m\cS^i_\m) + \ol C^a(\frac{i}{2}(\si^{-1}_1)_{ab}B^b +
M_a{}^i_\la p^\la_i), \eeq where $h^{pr}(x)$ is a non-degenerate
positive-definite matrix function on $X$ and $b^p{}_i^\m\cS^i_\m$
are the gauge-fixing combinations (\ref{y1}).

Thirdly, the desired gauge-fixing Lagrangian $\cL_{\rm GF}$ is
derived from the extended Lagrangian \be \cL'_{\rm PS}=\cL_{\rm
PS}+ i\ol C^{*p} B_p +i\ol C^*_a B^a, \ee where $\ol C^{*p}$, $\ol
C^*_a$ are antifields of auxiliary fields $\ol C_p$, $\ol C^a$ by
replacement of antifields with the variational derivatives
\mar{y12}\ben && y^*_i=\frac{\dl \Psi}{\dl y^i}, \qquad
C^*_p=\frac{\dl \Psi}{\dl C^p}=0, \qquad \ol
C^{*p}=\frac{\dl\Psi}{\dl \ol C_p}=
\frac{i}{2}  h^{pr}B_r +b^p{}_i^\m\cS^i_\m, \label{y12}\\
&& p^{*i}_\la =\frac{\dl\Psi}{\dl p^\la_i}=\ol C^a M_a{}^i_\la,
\qquad \ol C^*_a=\frac{\dl\Psi}{\dl \ol C^a}= \frac{i}{2}
(\si_1^{-1})_{ab}B^b + M_a{}^i_\la p^\la_i. \nonumber \een We
obtain \mar{y4}\beq \cL_{\rm GF}=\cL_\Pi+(\dl_i\Psi u^i_C
+\dl^i_\la\Psi u_{Ci}^\la) - B_p(\frac12  h^{pr}B_r
-ib^p{}_i^\m\cS^i_\m)  - B^a(\frac12(\si_1^{-1})_{ab} B^b -i
M_a{}^i_\la p^\la_i). \label{y4} \eeq Let us bring its second term
into the form \be && (\dr_i\Psi-d_\la\dr_i^\la\Psi) u^i_C
+\dr^i_\la\Psi u_{Ci}^\la= \dr_i\Psi u^i_C+ \dr_i^\la\Psi
d_\la(u^i_C)-
d_\la(\dr_i^\la\Psi u^i_C) +\dr^i_\la\Psi u_{Ci}^\la=\\
&& \qquad \ol u_C(\Psi)-d_\la(\dr_i^\la\Psi u^i_C), \ee where
\mar{y6}\beq \ol u_C=u^i_C\dr_i + u_{Ci}^\la\dr^i_\la +d_\la
u^i_C\dr^\la_i, \qquad d_\la=\dr_\la + y^i_\la\dr_i +C^r_\la
\frac{\dr}{\dr C^r} \label{y6} \eeq is the jet prolongation of the
vector field $u_C$ (\ref{y7}). In view of the transformation law
(\ref{m54}), we have \mar{y8}\ben && \ol u_C(\Psi)=-\ol
C_pb^p{}_i^\la J^1u_C(\cS^i_\la) -\ol C^a u_C(M_a{}^i_\la p^\la_i)
= -\ol C_pb^p{}_i^\la[\dr_\la u^i_rC^r + u^i_rC^r_\la +\dr_\la
u^{i\m}_r C^r_\m +u^{i\m}_r C^r_{\la\m}
+\nonumber\\
&& \qquad (\dr_ju^i_rC^r +\dr_ju^{i\m}_rC^r_\m)\cS^j_\la] -\ol C^a
M_a{}^i_\la ((M^{-1})^b{}_i^\la C_b -\dr_i u^k\cP^\la_k) = -\ol
C_p \cM^p_r C^r -\ol C^a C_a , \label{y8} \een where $\cM^p_r C^r$
is a second order differential operator on ghosts $C^r$. Then the
gauge-fixing Lagrangian (\ref{y4}) up to a divergence term takes
the form \mar{y9}\beq \cL_{\rm GF}=\cL_\Pi -\ol C_p \cM^p_r C^r
-\ol C^a C_a - \frac12 h^{pr}B_pB_r + iB_p b^p{}_i^\m\cS^i_\m
-\frac12 B^a(\si_1^{-1})_{ab} B^b +iB^aM_a{}^i_\la p^\la_i.
\label{y9} \eeq

Finally, one can write the generating functional \be &&
Z=N^{-1}\int \exp\{\int (\cL_\Pi -\ol C_p \cM^p_r C^r -\ol C^a C_a
- \frac12
h^{pr}B_pB_r +iB_p b^p{}_i^\m\cS^i_\m \\
&& \qquad -\frac12 B^a(\si_1^{-1})_{ab} B^b +iB^aM_a{}^i_\la
p^\la_i
+iJ_ky^k +iJ^i_\la p^\la_i)\om\}\\
&& \qquad  \op\prod_x [dB_p][dB^a][d\ol C_r][dC^r][d\ol C^a][dC_a]
[dy(x)][dp(x)] \ee of Euclidean Green functions. Integrating $Z$
as a Guassian integral with respect to the variables $B_p$, $B^a$
we obtain \mar{y10}\ben && Z=N'^{-1}\int \exp\{\int (\cL_\Pi  -\ol
C_p \cM^p_r C^r -\ol C^a C_a - \frac12 h_{pr}^{-1}
b^p{}_i^\m b^r{}_j^\nu\cS^i_\m \cS^j_\nu - \label{y10}\\
&& \qquad \frac12\si_1{}^{ab} M_a{}^i_\la M_b{}^j_\m p_i^\la
p_j^\m
 +iJ_ky^k+iJ^i_\la p^\la_i)\om\}\op\prod_x [d\ol C_r][dC^r][d\ol
C^a][dC_a] [dy(x)][dp(x)].\nonumber \een Of course, the Lagrangian
\mar{y11}\beq \cL_\Pi -\ol C_p M^p_r C^r -\ol C^a C_a - \frac12
h_{pr}^{-1} b^p{}_i^\m b^r{}_j^\nu\cS^i_\m \cS^j_\nu
-\frac12\si_1{}^{ab} M_a{}^i_\la M_b{}^j_\m p_i^\la p_j^\m
\label{y11} \eeq in the generating functional (\ref{y10}) is not
gauge-invariant, but it is invariant under the BRST transformation
\mar{m60}\ben && \vt=u^i_C\dr_i +u^\la_{Ci}\dr^i_\la +d_\la
u^i_C\dr_i^\la +
\ol v^a\frac{\dr}{\dr \ol C^a} + v_a\frac{\dr}{\dr C_a} + \label{m60}\\
&& \qquad \ol v_r\frac{\dr}{\dr\ol C_r} +
 v^r \frac{\dr}{\dr C^r}  +d_\la v^r \frac{\dr}{\dr C^r_\la}
+d_\m d_\la v^r \frac{\dr}{\dr C^r_{\m\la}},  \nonumber\\
&& d_\la=\dr_\la +y^i_\la\dr_i + y^i_{\la\m}\dr_i^\m + C^r_\la
\frac{\dr}{\dr C^r} +C_{\la\m}^r\frac{\dr}{\dr C^r_\m}, \nonumber
\een whose components $v$ are given by the antibrackets \be &&
v_a=(C_a, \cL'_{\rm PS})=\frac{\dl \cL'_{\rm PS}}{\dl C^{*a}}=0,
\qquad \ol v^a= (\ol C^a, \cL'_{\rm PS})=\frac{\dl \cL'_{\rm
PS}}{\dl
\ol C^*_a}= iB^a\\
&& v^r=(C^r,\cL'_{\rm PS})=\frac{\dl \cL'_{\rm PS}}{\dl
C^*_r}=-\frac12 c^r_{pq} C^pC^q, \qquad
 \ol v_r=(\ol C_r,\cL'_{\rm PS})=\frac{\dl \cL'_{\rm PS}}{\dl \ol C^{*r}}=iB_r
\ee restricted to the shell (\ref{y12}) and to the solutions \be
B_r=ih^{-1}_{rp} b^p{}_i^\m\cS^i_\m, \qquad
B^a=i\si_1{}^{ab}M_b{}^i_\la p^\la_i \ee of the Euler--Lagrange
equations $\dl\cL_{\rm GF}/\dl B_r=0$ and $\dl\cL_{\rm GF}/\dl
B^a=0$.

Integration of the generating functional $Z$ (\ref{y10}) as a
Gaussian integral with respect to the variables $\ol C^a$, $C_a$
results in the desired BV quantization \mar{y107}\ben &&
Z=N'^{-1}\int \exp\{\int (\cL_\Pi -\ol C_p \cM^p_r C^r  - \frac12
h_{pr}^{-1}
b^p{}_i^\m b^r{}_j^\nu\cS^i_\m \cS^j_\nu - \label{y107}\\
&& \qquad \frac12\si_1{}^{ij}_{\la\m} p_i^\la p_j^\m
 +iJ_ky^k+iJ^i_\la p^\la_i)\om\}\op\prod_x [d\ol
C_r][dC^r][dy(x)][dp(x)].\nonumber \een of the Lagrangian $L_\Pi$
(\ref{m32}). Integrating the generating functional $Z$
(\ref{y107}) as a Gaussian integral with respect to the momenta
$p(x)$ under the condition $J^i_\la=0$, one restarts the
generating functional \be && Z=N'^{-1}\int \exp\{\int (\cL -\ol
C_p M^p_r C^r - \frac12 h_{pr}^{-1} b^p{}_i^\m b^r{}_j^\nu\cS^i_\m
\cS^j_\nu +iJ_ky^k)\om\}\op \prod_x [d\ol C_r][dC^r][dy(x)] \ee of
the BV quantization of the original quadratic Lagrangian $L$
(\ref{N12}).

\section{Hamiltonian gauge theory}

For example, let us consider gauge theory of principal connections
on a principal bundle $P\to X$ with a structure Lie group $G$.
Principal connections on $P\to X$ are represented by sections of
the affine bundle \mar{br3}\beq C=J^1P/G\to X, \label{br3} \eeq
modelled over the vector bundle $T^*X\ot V_GP$ \cite{book}. Here,
$V_GP=VP/G$ is the fiber bundle in Lie algebras $\cG$ of the group
$G$. Given the basis $\{\ve_r\}$ for $\cG$, we obtain the local
fiber bases $\{e_r\}$ for $V_GP$. The connection bundle $C$
(\ref{br3}) is coordinated by $(x^\m,a^r_\m)$ such that, written
relative to these coordinates, sections $A=A^r_\m dx^\m\ot e_r$ of
$C\to X$ are the familiar local connection one-forms, regarded as
gauge potentials.

There is one-to-one correspondence between the sections $\xi=\xi^r
e_r$ of $V_GP\to X$ and the vector fields on $P$ which are
infinitesimal generators of one-parameter groups of vertical
automorphisms (gauge transformations) of $P$. Any section $\xi$ of
$V_GP\to X$ yields the vector field \mar{br6}\beq u(\xi)=u^r_\m
\frac{\dr}{\dr a^r_\m}=(c^r_{pq}a^p_\m\xi^q+ \dr_\m\xi^r)
\frac{\dr}{\dr a^r_\m} \label{br6} \eeq on $C$, where $c^r_{pq}$
are the structure constants of the Lie algebra $\cG$.

The configuration space of gauge theory is the jet manifold $J^1C$
equipped with the coordinates $(x^\la,a^r_\la,a^r_{\m\la})$. It
admits the canonical splitting (\ref{N18}) given by the coordinate
expression \mar{N31'}\beq a^r_{\m\la}=\cS^r_{\m\la}+\cF^r_{\m\la}=
\frac12(a^r_{\m\la}+a^r_{\la\m}-c^r_{pq}a^p_\m a^q_\la)
+\frac12(a^r_{\m\la}-a^r_{\la\m} +c^r_{pq}a^p_\m a^q_\la), \eeq
where $\cF$  is the strength of gauge fields up to the factor 1/2.
The Yang--Mills Lagrangian on the configuration space $J^1C$ reads
\mar{5.1'}\beq L_{\rm YM}=a^G_{pq}g^{\la\m}g^{\bt\n}\cF^p_{\la
\beta}\cF^q_{\m\n}\sqrt{\nm g}\,\om, \qquad  g=\det(g_{\m\nu}),
\label{5.1'} \eeq where  $a^G$ is a non-degenerate $G$-invariant
metric in the dual of the Lie algebra of ${\got g}$ and $g$ is a
nondegenerate metric on $X$.

The phase space $\Pi$ (\ref{00}) of the gauge theory is endowed
with the canonical coordinates $(x^\la,a^p_\la,p^{\mu\la}_q)$. It
admits the canonical splitting (\ref{N20}) given by the coordinate
expression \mar{N32}\beq p^{\mu\la}_m= \cR^{\mu\la}_m +
\cP^{\mu\la}_m= p^{(\mu\la)}_m +
p^{[\mu\la]}_m=\frac{1}{2}(p^{\mu\la}_m+ p^{\la\mu}_m) +
\frac{1}{2}(p^{\mu\la}_m- p^{\la\mu}_m). \label{N32} \eeq With
respect to this splitting, the Legendre map induced by the
Lagrangian (\ref{5.1'}) takes the form \mar{5.2a,b} \ben
 &&p^{(\mu\la)}_m\circ\wh L_{YM}=0, \label{5.2a}\\
&&p^{[\mu\la]}_m\circ\wh L_{YM}=4a^G_{mn}g^{\m\al}g^{\la\bt}
\cF^n_{\al\bt}\sqrt{|g|}. \label{5.2b} \een The equalities
(\ref{5.2a}) define the Lagrangian constraint space $N_L$ of
Hamiltonian gauge theory. Obviously, it is an imbedded submanifold
of $\Pi$, and the Lagrangian $L_{\rm YM}$ is almost-regular.

In order to construct an associated Hamiltonian, let us consider a
connection $\G$ (\ref{N16}) on the fiber bundle $C\to X$ which
take their values into $\Ker\wh L$, i.e., \be
\G^r_{\la\m}-\G^r_{\m\la}+c^r_{pq}a^p_\la a^q_\m=0. \ee Given a
symmetric linear connection $K$ on $X$ and a principal connection
$B$ on $P\to X$, this connection reads \be
\G^r_{\la\m}=\frac{1}{2} [\dr_\mu B^r_\la+\dr_\la B^r_\mu
-c^r_{pq}a^p_\la a^q_\mu  +  c^r_{pq} (a^p_\la B^q_\m +a^p_\m
B^q_\la)] - K_\la{}^\bt{}_\mu(a^r_\bt-B^r_\bt). \ee The
corresponding Hamiltonian (\ref{N22}) associated to $L_{\rm YM}$
is \be \cH_\G=p^{\la\m}_r\G^r_{\la\m}+a^{mn}_Gg_{\mu\nu}
g_{\la\beta} p^{[\mu\la]}_m p^{[\nu\bt]}_n\sqrt{|g|}. \ee Then we
obtain the Lagrangian \be \cL_N=p^{[\la\m]}_r\cF^r_{\la\m}-
a^{mn}_Gg_{\mu\nu} g_{\la\beta} p^{[\mu\la]}_m
p^{[\nu\bt]}_n\sqrt{|g|} \ee (\ref{bv2}) on $N_L\op\times_Y J^1Y$
and its pull-back \mar{m63}\beq L_\Pi=\cL_\Pi\om, \qquad
\cL_\Pi=p^{\la\m}_r\cF^r_{\la\m}- a^{mn}_Gg_{\mu\nu} g_{\la\beta}
p^{[\mu\la]}_m p^{[\nu\bt]}_n\sqrt{|g|}, \label{m63} \eeq
(\ref{m32}) onto $\Pi\op\times_Y J^1Y$.

Both the Lagrangian $L_{\rm YM}$ (\ref{5.1'})  and the Lagrangian
$L_\Pi$ (\ref{m63}) are invariant under gauge transformations
whose infinitesimal generators are the lifts \be &&
J^1u(\xi)=(c^r_{pq}a^p_\m\xi^q+\dr_\m\xi^r) \frac{\dr}{\dr a^r_\m}
+ (c^r_{pq}(a^p_{\la\m}\xi^q +a^p_\m\dr_\la\xi^q)
+\dr_\la\dr_\m\xi^r)\frac{\dr}{\dr a^r_{\la\m}},\\
&& \ol u(\xi)=J^1u(\xi) - c^r_{pq}p_r^{\la\m}\xi^q \frac{\dr}{\dr
p_p^{\la\m}} \ee of the vector fields (\ref{br6}) onto $J^1C$ and
$\Pi\op\times_C J^1C$, respectively. We have the transformation
laws \be J^1u(\xi)(\cF^r_{\la\m})=c^r_{pq}\cF^p_{\la\m}\xi^q,
\qquad J^1u(\xi)(\cS^r_{\la\m})= c^r_{pq}\cS^p_{\la\m}\xi^q
+c^r_{pq}a^p_\m\dr_\la\xi^q+ \dr_\la\dr_\m\xi^r. \ee Therefore,
one can choose the gauge conditions \be
g^{\la\m}S^r_{\la\m}(x)-\al^r(x)= \frac12g^{\la\m}(\dr_\la
a^r_\m(x) +\dr_\m a^r_\la(x))-\al^r(x)=0, \ee which are the
familiar generalized Lorentz gauge. The corresponding second-order
differential operator (\ref{y8}) reads \be
\cM^r_s\xi^s=g^{\la\m}(\frac12 c^r_{pq} (\dr_\la a^p_\m +\dr_\m
a^p_\la)\xi^q +c^r_{pq}a^p_\m\dr_\la\xi^q +\dr_\la\dr_\m\xi^r).
\ee Passing to the Euclidean space and repeating the quantization
procedure in previous Section, we come to the generating
functional \be && Z=\cN^{-1}\int\exp\{\int
(p^{\la\m}_r\cF^r_{\la\m}- a^{mn}_Gg_{\mu\nu}
g_{\la\beta} p^{\mu\la}_m p^{\nu\bt}_n\sqrt{|g|}-\\
&& \qquad \frac18 a^G_{rs}g^{\al\nu}g^{\la\m}(\dr_\al a^r_\nu
+\dr_\nu a^r_\al) (\dr_\la a^s_\m +\dr_\m a^s_\la) -g^{\la\m}\ol
C_r(\frac12 c^r_{pq} (\dr_\la a^p_\m
+\dr_\m a^p_\la)C^q+c^r_{pq}a^p_\m C^q_\la +C_{\la\m}^r)  \\
&& \qquad + iJ_r^\mu a^r_\mu +iJ^r_{\m\la} p_r^{\m\la})\om\}
\op\prod_x [d\ol C][d C][dp(x)][da(x)]. \ee (\ref{y107}) of BV
quantization of Hamiltonian gauge theory. Its integration with
respect to momenta restarts the familiar BV quantization of gauge
theory.

\end{document}